# RD51
## An R&D collaboration for micropattern gaseous detectors

Serge Duarte Pinto*
on behalf of the RD51 collaboration

July 16, 2009

Abstract — The RD51 collaboration was founded in April 2008 to coordinate and facilitate efforts for development of micropattern gaseous detectors (MPGDs). The 59 institutes from 20 countries bundle their effort, experience and resources to develop these emerging micropattern technologies.

MPGDs are already employed in several nuclear and high-energy physics experiments, medical imaging instruments and photodetection applications; many more applications are foreseen. They outperform traditional wire chambers in terms of rate capability, time and position resolution, granularity, stability and radiation hardness. RD51 supports efforts to make MPGDs also suitable for large areas, increase cost-efficiency, develop portable detectors and improve ease-of-use.

The collaboration is organized in working groups which develop detectors with new geometries, study and simulate their properties, and design optimized electronics. Among the common supported projects are creation of test infrastructure such as beam test and irradiation facilities, and the production workshop.

## Introduction

The working principle of all gaseous radiation detectors is similar: radiation generates primary ionization in the gas, and the primary electrons create further electron-ion pairs in an avalanche process in a region with a strong electrostatic field. Gaseous detectors differ in how this strong field region is created. For several decades the most popular way was using thin wires, either one or many, where close to the wire the field strength is inversely proportional to the distance to the wire. This is illustrated in Fig. 1, the two first pictures.

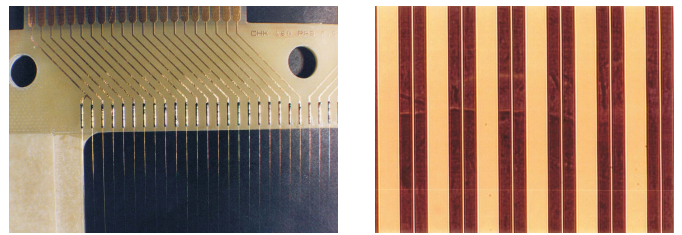

Figure 2 — Left: wires of a multiwire proportional chamber (MWPC) soldered to a frame. Right: microscope image of a microstrip gas chamber (MSGC)

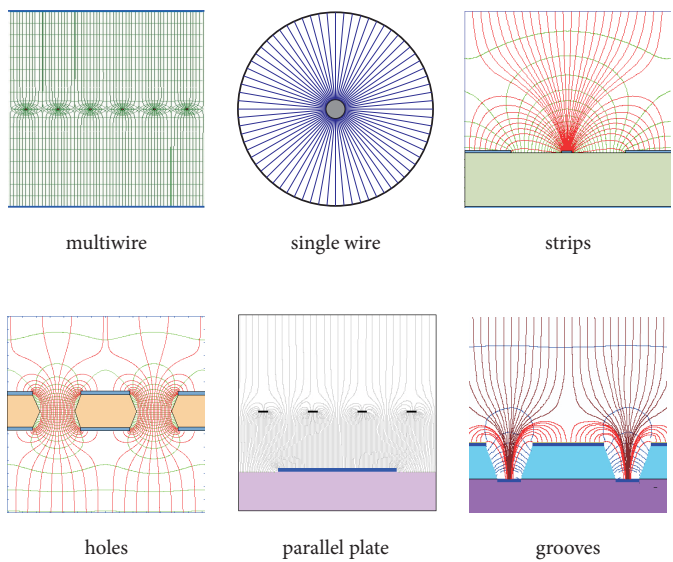

Figure 1 — Various technologies

In recent years, many planar structures have emerged that generate an enhanced field region in various ways. Several examples are shown in Fig. 1 and still many more have been developed. Common feature among all these structures is a narrow amplification gap of typically 50–100 microns, compared to many millimeters for wire-type structures. These devices are now known under the common name of micropattern gaseous detectors (MPGDs).

### Microstrip gas chamber

The first such structure to gain popularity was the microstrip gas chamber [1] (MSGC), of which the field pattern is shown in the third picture in Fig. 1. The principle of an MSGC resembles a wire chamber, with fine printed strips instead of thin wires, see Fig. 2. However, the spacing between anode strips was as narrow as 200 microns (compared to at least several millimeters for wire chambers) due to the microelectronics techniques employed in manufacturing. Most ions created in the avalanche process drift to the wider cathode strips, which are spaced only 60 microns away from the anodes. This short drift path for ions overcomes the space charge effect present in wire chambers, where the slowly drifting ions may remain in the gas volume for milliseconds, and modify the electric field (thereby reducing the gain). Figure 3 shows how this space charge effect limits the rate capability of wire chambers, and how the fine granularity of MSGCs pushes this limit by two orders of magnitude.

The high rate capability of the MSGC made it an attractive technology for many applications. However, the development of the MSGC also indicated some new limitations, most of which are common to all micropattern devices. Possibly most important is the issue of discharges, which eventually led high-energy experiments to abandon MSGC technology. Another issue largely common to micropattern gas detectors is the charging of insulating surfaces which modifies the field shape locally, limiting the time stability. For MSGCs this could be solved by surface treatment of the glass substrate to decrease the surface resistivity.

*Serge.Duarte.Pinto@cern.ch

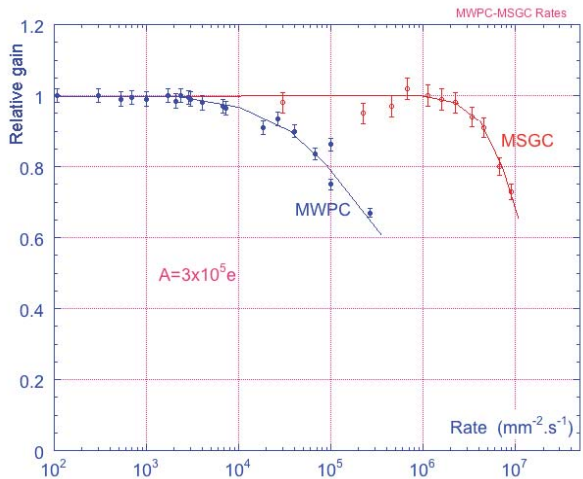

Figure 3 — Gain as a function of particle rate in otherwise constant conditions, for wire chambers in blue and MSGCs in red.

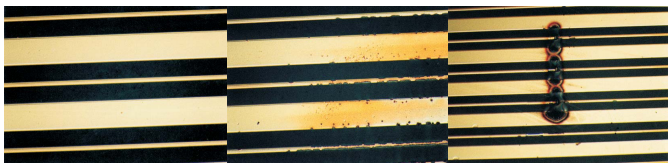

Figure 4 — Damage done to MSGCs by discharges. In the rightmost frame anode strips are cut, leaving part of those anodes inactive. With its very thin metal layers MSGCs are particularly vulnerable for discharge damage.

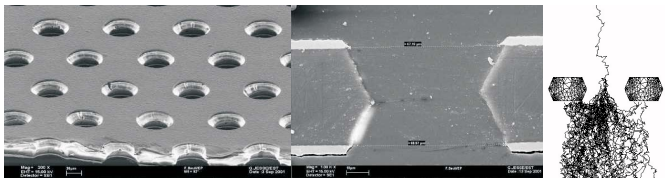

Figure 5 — Electron microscope images of a GEM foil, and a simulated electron avalanche in a GEM hole.

The microstrip gas chamber suffered severely from discharges (induced by heavily ionizing particles), which could fatally damage the fragile anode strips, see Fig. 4. In 1997 the gas electron multiplier (GEM) was introduced [2] as a preamplification stage for the MSGC. This allowed the MSGC to work at a lower voltage, thereby lowering the probability of discharges as well as the energy involved in discharges when they occurred. The GEM principle was so successful that it soon became the basis for a detector in its own right.

*Gas electron multiplier*

The gas electron multiplier is a copper clad polyimide foil with a regular pattern of densely spaced holes, see Fig. 5. Upon applying a voltage between top and bottom electrodes, a dipole field is formed which focuses inside the holes where it is strong enough for gas amplification. As a GEM is only an amplification structure it is independent of the readout structure, which can be optimized for the application (see a few examples in Fig. 6). Due to the separation from the readout structure, possible discharges do not directly impact the front-end electronics, thus making the detector more discharge tolerant. Also, it can be cascaded to achieve higher gain at lower GEM voltage, which decreases the discharge probability, see Fig. 7. The triple GEM has now become a standard which is used in many high rate applications [4, 5, 6].

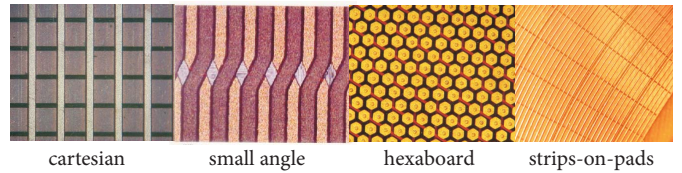

Figure 6 — Some examples of readout structures developed for GEM detectors.

*Micromegas*

Another detector structure developed about the same time is the micromesh gas detector, or Micromegas [7]. This detector has a parallel plate geometry with the amplification gap between a micromesh and the readout board. Parallel plate amplification existed before, but the Micromegas has a much narrower amplification gap of around 50–100 $\mu$m. The narrow amplification gap provides fast signals and a high rate capability. The micromesh is supported by regularly spaced pillars which maintain the accurate spacing. This is shown in Fig. 8.

### Current trends in MPGDs

The development of MPGDs took off in the 1990s mainly as a way to achieve a higher rate capability with gaseous detectors. Since then applications have driven developers to exploit the additional benefits of these structures, such as excellent time and position resolution, resistance to aging, and intrinsic ion and photon feedback suppression. Advances in available techniques for microelectronics and printed circuits opened ways to make new structures and optimize existing ones. This led to a wide range of detector structures for an even wider range of applications, with a performance superior to any traditional gas detector.

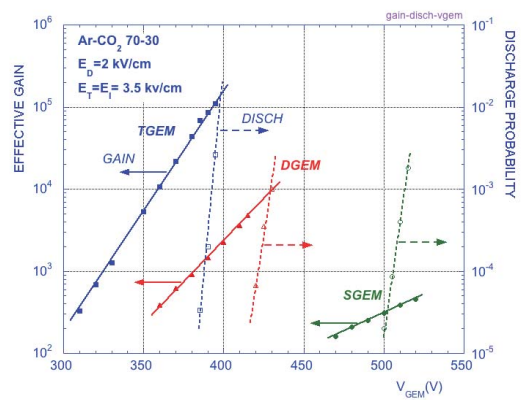

Figure 7 — Gain and discharge probability as a function of GEM voltage, for single, double and triple GEM detectors. Discharge probability is measured by irradiation with $\alpha$-particles, which are so strongly ionizing that they are likely to cause a discharge.

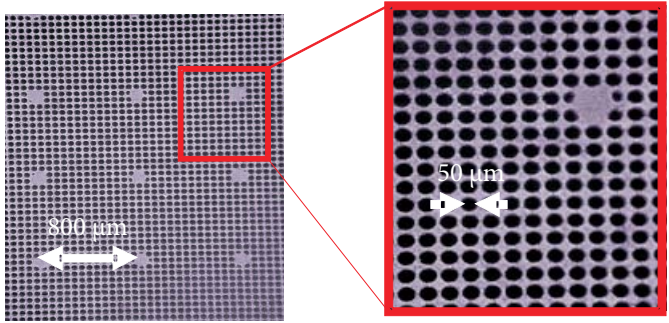

Figure 8 — Microscope images of a Micromegas detector, with indicated mesh and pillar spacings.

*Techniques*

The techniques that enabled the advent of micropattern gas detectors come from the industry of microelectronics and printed circuits. The microstrip gas chamber was made by employing photolithographic techniques used by microelectronics manufacturers. Instead of silicon wafers, thin glass plates were used as a substrate for printing the fine strip patterns. These glass plates were doped or sputter coated with so-called Pestov glass in order to reduce slightly the surface resistivity, which improved the time stability [8].

The very thin metal layers of MSGCs (few hundred nanometers) makes them vulnerable for discharges, which can easily do fatal damage (Fig. 4). Many of the later micropattern devices use thicker metals (few microns), and performance is normally unaffected by thousands of discharges. The techniques used to pattern these metals and the insulators separating them come from the manufacturing of printed circuit boards (PCBs). An advantage is the much lower cost, and the possibility to cover large areas. These techniques include photolithography, metal etching and screen printing.

A rather special technique, thoroughly refined in the CERN PCB workshop, is the etching of polyimide. This is the basis of a number of micropattern gas detectors, including the GEM. Another more industry standard method to pattern insulators is using photo-imageable polymers, such as photoresist, coverlayers and solder masks.

*Technologies*

A few of the most prominent micropattern gas detector technologies have been mentioned in the introduction. Many more types of structures were developed and are currently used, which are often derived from MSGC, GEM or Micromegas. A few more examples are discussed here, but the selection is by no means exhaustive.

The refinement of the polyimide etching technique that is used to make GEMs, is also used for some detectors with a readout structure in the same plane as the amplification structure. These are the WELL [9] and the *groove detector* [10]. Unlike the GEM these structures are not "transparent", all the electrons from the avalanche are collected on the bottom electrode which is also the readout structure. The *microhole and strip plate* [11] combines the amplification mechanisms of GEM holes and microstrips (see Fig. 9, first frame), and combines a high gas gain with an unparalleled ion feedback suppression.

Another GEM-derivative is the *thickGEM* [12], also shown in Fig. 9. This is a hole-type amplification structure, where the flexible polyimide substrate is replaced by a thicker glassfiber-reinforced-epoxy plate and the holes are mechanically drilled. The substrate is the standard base material for rigid PCBs and is therefore cheap, and readily available from any PCB manufacturer. Also the automatic drilling of the holes is a standard industry procedure. One has full control over the hole pitch and diameter, and the shape, size and thickness of the base material. These structures are convenient for applications where position and time resolution are not the most critical parameters, but which require a high gain and a certain ruggedness. ThickGEMs are for instance popular for photodetector applications, where the stiff substrate lends itself well to the vacuum deposition of a CsI photoconverter [13]. More recently, electrodes of thickGEMs have been covered with or replaced by resistive layers [14]. These detectors are reported to work stably in streamer mode, due to the enhanced quenching by the resistive layers.

Micromegas detectors underwent a technical improvement with the introduction of a new fabrication method [15]. Here a woven metal micromesh is laminated to the readout board between layers of photoimageable soldermask. These soldermask layers can subsequently be patterned by UV-exposure to create the supporting pillar structure (see the third frame of Fig. 9). The materials involved are quite inexpensive, and the processes are industry standard, which makes it suitable for large scale production. Also, the homogeneity of the grid spacing is better than of the original Micromegas detectors, and the detector is very robust.

The *micropin array* [16] was introduced for x-ray imaging (see Fig. 9, fourth frame). The spherical geometry of the electric field close to the end of each pin (proportional to $1/r^2$ compared to $1/r$ of a wire chamber) gives rise to very short amplification region, allowing a rate-stable high gain. A similar philosophy led to the development of the *microdot chamber* [17], for which microelectronics techniques were employed to reach feature sizes of only a few microns.

The coming of age of *post-wafer processing* techniques marked the introduction of MPGDs with pixel readout. These detectors use the bump-bonding pads of a pixel chip as a readout structure. The position and time resolution of these devices is unmatched by any other gas detector. Due to their high sensitivity they can distinguish each primary electron. This enables them to resolve delta-rays from a track or to reconstruct the direction of emission of a photoelectron from an x-ray conversion (related to the x-ray polarization). One group uses a Micromegas-type of gas amplification: *InGrid* [19]. The grid electrode and the insulating pillar structure supporting it are made directly on the chip by post-wafer processing techniques, allowing the grid holes to be aligned with the readout pads (see Fig. 9, fifth frame). Another group uses an ASIC with a hexagonal readout pad structure, and a GEM-based amplification structure [18] Here the GEM has a reduced pitch of $50\mu m$ and thickness of $25\mu m$ (compared to $140\mu m$ and $50\mu m$ respectively for standard GEMs) to match the granularity of the readout (see the last frame of Fig. 9).

*Applications*

Micropattern gas detectors have already been applied in many instruments and experiments, both by science and industry. Possible fields of application are high-energy and nuclear physics, synchrotron and thermal neutron research, medical imaging and homeland security. Most structures were primarily developed for high rate tracking of charged particles in nuclear and high-energy physics experiments. For instance Micromegas [20] and GEMs [4] are used in the COMPASS experiment, and GEMs in LHCb [6] and TOTEM [5] experiments. Also for the LHC machine upgrade program to increase its luminosity by roughly a factor of ten, most of the experiments foresee replacement of wire chambers, drift tubes and resistive plate chambers by MPGDs. However many MPGDs have shown to be suitable for other applications as well. A few examples are given here.

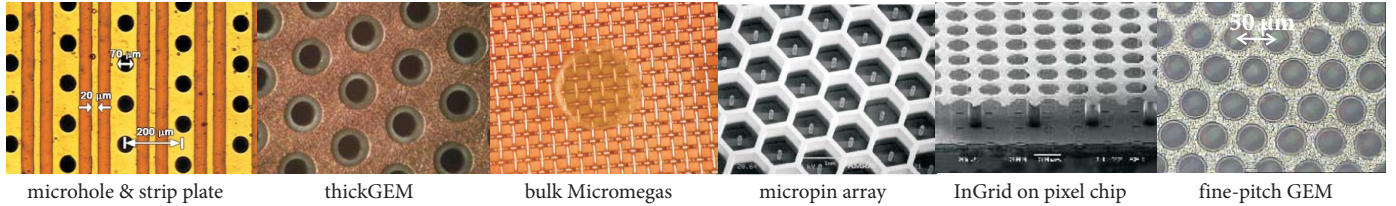

| microhole & strip plate | thickGEM | bulk Micromegas | micropin array | InGrid on pixel chip | fine-pitch GEM |

Figure 9 — Microscope images of various detector structures. See text for details on each frame.

Both GEMs and Micromegas can be used for the readout of a time projection chamber [21] (TPC). Compared to wire chambers, these MPGDs have the benefit that the planar structure suppresses the so-called $E \times B$ effects which limit the spatial resolution of wire chambers in TPC configuration. Also, both Micromegas and GEMs have a natural ion feedback suppression, which may make a gating structure unnecessary.

As mentioned before, GEM-like structures can be coated with a photoconverter (typically CsI) to serve as a photon counter. In this way, large areas can be covered with hardly any dead zones, and the technique is inexpensive. This makes it attractive for ring imaging Cherenkov detectors, of which the photodetector planes often span several square meters. Also here the ion feedback suppression is an added benefit, as it increases the lifetime of the photoconverter. In addition, the detector can be made "hadron-blind" by reversing the drift field, and even "windowless" if the Cherenkov radiator gas (in that case typically $CF_4$) is also used as amplification gas [22].

X-ray counting and imaging detectors can be based on MPGDs [23], as x-rays convert in some noble gases leaving typically few hundred primary electrons for detection. For these purposes efficient x-ray conversion gases are frequently used, such as xenon or krypton. Argon is about an order of magnitude less efficient, but so much cheaper that it can still be attractive for high rate applications.

Microstrip gas chambers and GEM detectors are used as neutron detectors [24]. Typically a boron layer (in the form of $B_2O_3$) is evaporated onto the GEM foils, which acts as a neutron converter via the reaction $^{10}B + n \rightarrow {}^7Li + \alpha$. In the case of MSGCs, $^3$He is often used as both amplification and convertor gas. Here the conversion reaction is: $^3He + n \rightarrow {}^3H + p$.

*Performance*

Depending on the application, the performance of MPGDs has different figures of merit. The first MPGDs were designed to obtain a high rate capability. Several MHz/mm² of charged particles are easily reached with, for instance, a triple GEM detector, without a measurable loss of gain and with negligible discharge probability.

Time, position and energy resolution are crucial figures for most applications. GEM-based detectors normally have a position resolution of about 50$\mu$m, Micromegas can go down to ~ 12$\mu$m if equipped with a high density readout board. Time resolutions are of the order of few nanoseconds. X-ray energy resolution is often measured using a $^{55}$Fe source, obtaining a FWHM between 15% and 22%. MPGDs with pixel chip readout report position resolutions below 10$\mu$m and a time resolution of 1 ns. From the $^{55}$Fe spectrum they can resolve the K$_\alpha$ and K$_\beta$ energies, and reach a resolution of 12%.

The reduction of ion backflow into the drift region is a general property of MPGDs. It is usually expressed as a fraction of the effective gain, and this value depends quite strongly on the way the fields are configured in the chamber. Microhole and strip plates feature a particularly effective ion feedback suppression of the order $10^{-4}$ in optimized conditions.

Aging modes of gas detectors are largely understood in the case of wire chambers. There the plasmas that are formed during avalanches in the strong field near the wire deposit layers of silica or polymers which reduce the gain and give rise to micro discharges. Most micropattern devices do not generate such a strong field at the surface of the conductors, and consequently little signs of aging have been observed. Aging studies of MPGDs specifically have rarely been done yet, and time will prove if they are as resistant to aging as it seems.

### An R&D collaboration for MPGDs

RD51 is a large R&D collaboration, which unites many institutes in an effort to advance technological development of micropattern gas detectors. At the time of writing there are ~ 350 participating authors from 59 institutes in 20 countries worldwide. The efforts of the collaboration do not focus on one or a few particular applications for MPGDs, but is rather *technology oriented*. It is a platform for sharing of information, results and experiences, and for steering R&D efforts. It tries to optimize the cost of R&D projects by sharing resources, creating common projects and providing common infrastructure.

*Organization*

RD51 has two spokespersons, Leszek Ropelewski[1] and Maxim Titov[2], who can be contacted for more information. Concerning all scientific matters the collaboration is governed by a *collaboration board* (CB), which is also responsible for coordinating the financial planning and other resource issues, in particular for managing the common fund. Representatives from all collaborating institutes are seated in the CB, and have voting rights. A *management board* (MB) supervises the progress of the work program along the lines defined by the CB and prepares decisions for and makes recommendations to the CB.

---

[1] Leszek.Ropelewski@cern.ch
[2] maxim.titov@cea.fr

Table 1 — Organization of RD51 in working groups and tasks.

| | WG1<br>MPGD technology<br>& new structures | WG2<br>Characterization | WG3<br>Applications | WG4<br>Software<br>& simulation | WG5<br>Electronics | WG6<br>Production | WG7<br>Common test<br>facilities |
|---|---|---|---|---|---|---|---|
| OBJECTIVES | Design optimization<br><br>Development of new geometries and techniques | Common test standards<br><br>Characterization and understanding of physical phenomena in MPGDs | Evaluation and optimization for specific applications | Development of common software and documentation for MPGD applications | Readout electronics optimization and intergration with MPGDs | Development of cost-effective technologies and industrialization | Sharing of common infrastructure for detector characterization |
| TASKS | Large area MPGDs<br>—<br>Design optimization<br>New geometries<br>Fabrication<br>—<br>Development of rad-hard detectors<br>—<br>Development of portable detectors | Common test standards<br>—<br>Discharge protection<br>—<br>Aging and radiation hardness<br>—<br>Charging-up and rate capability<br>—<br>Avalanche statistics | Tracking and triggering<br>—<br>Photodetection<br>—<br>Calorimetry<br>—<br>Cryogenic det.<br>—<br>X-ray & neutron imaging<br>—<br>Astroparticle physics appl.<br>—<br>Medical appl.<br>—<br>Synchrotron rad.<br>Plasma diagn.<br>Homeland sec. | Algorithms<br>—<br>Simulation improvements<br>—<br>Common platforms (ROOT, GEANT)<br>—<br>Electronics modeling | FE electronics requirements definition<br>—<br>General purpose pixel chip<br>—<br>Large area systems with pixel readout<br>—<br>Portable multi-channel system<br>—<br>Discharge protection strategies | Common production facility<br>—<br>industrialization<br>—<br>Collaboration with industrial partners | Testbeam facility<br>—<br>Irradiation facility |

The activity is divided in seven working groups (WGs), covering all relevant topics of MPGD-related R&D. A number of tasks is assigned to each working group. Table 1 lists all the WGs and indicates their objectives and tasks.

WG1 is concerned with the technology of MPGDs and the design of new structures. Examples are efforts to make Micromegas, GEM and thickGEM technologies suitable for large areas [25]. Also interesting is the development of cylindrical GEM [26] and Micromegas [27] detectors for inner barrel tracking. A recent development is the introduction of spherical GEMs [28] for parallax-free x-ray diffraction measurements.

The second working group deals with physics issues of MPGDs, such as discharges, charging of dielectric surfaces and aging. Also, common test standards are proposed to enable different groups to compare their results. Regular meetings have become a forum for exchanging results and for discussion about what are actually the most fundamental properties of micropattern gas detectors.

WG3 concentrates on the applications of MPGDs, and on how to optimize detectors for particularly demanding applications. Examples have been listed above and new applications still appear. Sometimes from surprising fields: one project aims to construct very large area GEM chambers to detect nuclear fission materials or waste in cargo containers by tomography of cosmic ray muons [29].

WG4 develops simulation software and makes progress in the field of simulation. Simulation is essential to understand the behavior of detectors. A mature range of software tools is available for simulating primary ionization (Heed[3]), electron transport properties in gas mixtures in electric and magnetic fields (Magboltz[4]), and gas avalanches and induction of signals on readout electrodes (Garfield[5]). Garfield has interfaces to Heed and Magboltz and only needs to be supplied with a field map and detector configuration. A field map can be generated by commercial finite-element method (FEM) programs such as Ansys, Maxwell, Tosca, QuickField and Femlab. Within the collaboration, an open-source field solver is developed and recently released called neBEM [30]. It is based on the boundary element method (BEM), and is in most respects superior to FEM solvers for gas detector simulations.

Front-end electronics and data acquisition systems are discussed in WG5. Electronics for detectors are highly specialized and therefore almost entirely based on application specific integrated circuits (ASICs). A front-end ASIC often has to be radiation tolerant, must accept external triggers and have long analog pipelines for the trigger latency, and must support high output data rates. Availability, flexibility and scalability of chips and DAQ systems are discussed in regular meetings. MPGDs have typically one more requirement for the front-end chip: it must survive discharges, and the dead time following a discharge must be kept to a minimum. Various solutions are in development in this working group.

WG6 deals with the production of MPGDs. Almost all MPGDs were first made in the CERN PCB workshop of Rui de Oliveira,

---
[3] Author: Igor Smirnov (http://consult.cern.ch/writeup/heed/)
[4] Author: Stephen Biagi (http://consult.cern.ch/writeup/magboltz/)
[5] Author: Rob Veenhof (http://garfield.web.cern.ch/garfield/)

and it remains an almost exclusive manufacturing site for most technologies. Hence, efforts in WG6 are aimed at plans for upgrading this workshop on one hand, and industrial partnership and export of the technology and know-how on the other. Also, scenarios are developed for industrial scale production of some MPGDs (especially GEMs and Micromegas), in case a large experiment decides to implement them in their system.

Finally, WG7 coordinates the effort to set up a shared test infrastructure in the form of test beam and irradiation facilities. The test beam facility will be equipped with supply and exhaust of gases, including flammable mixtures. Also a large 1.4 Tesla magnet will be provided. The irradiation facility provides a strong gamma source (a 10 TBq $^{137}$Cs source is foreseen) combined with a 100 GeV muon test beam ($10^4$ muons per spill) and is called GIF++ [31].

## Conclusions and contacts

Micropattern gas detectors have a great potential in science and industry, in medical and commercial applications. RD51 is committed to fulfill this potential. The collaboration welcomes new institutes who are interested in participating in the development of micropattern gas detectors. Up-to-date information and relevant contacts can be found on the collaboration webpage[6], or by contacting the spokespersons.

---

[6] http://rd51-public.web.cern.ch/RD51-Public/